\documentstyle[12pt]{article}
\begin{document}
\newcommand{\be}{\begin{equation}}
\newcommand{\ee}{\end{equation}}
\newcommand{\ba}{\begin{array}}
\newcommand{\ea}{\end{array}}
\newcommand{\bea}{\begin{eqnarray}}
\newcommand{\eea}{\end{eqnarray}}
\def\vphi{\varphi}
\def\vrho{\varrho}
\def\vtheta{\vartheta}
\begin{center}
{\large{\bf Two exactly-solvable problems

\vspace{0.2cm}
in one-dimensional hyperbolic Space}}

\vspace{0.5cm}
{\bf \v{C}. Burd\'{\i}k$^{1,2}$ and G.S.Pogosyan$^{1,3}$}

\vspace{0.3cm}
$^{1}${\it Laboratory of Theoretical Physics, Joint Institute for Nuclear
Research, \\
Dubna, 141980, Moscow Region, Russia}

\vspace{0.3cm}
$^{2}$
{\it Department of Mathematics Czech Technical University,
Trojanova 13, 120 00, Prague, Czech Republic}

\vspace{0.3cm}
$^{3}${\it Centro de Ciencias F\'{\i}sicas
UNAM, Apartado Postal 48--3 \\
62251 Cuernavaca, Morelos, M\'exico}

\end{center}
\vspace{0.6cm}

\begin{abstract}
In this paper we establish a relation between two exactly-solvable
problems on one-dimensional hyperbolics space, namely singular
Coulomb and singular oscillator systems.

\end{abstract}

\vspace{0.6cm}

\noindent
{\bf 1.}
In the last few years the relation between Coulomb and oscillator
problems on the N-dimensional spaces with constant curvature
(both positive on the sphere $S_n$ and negative on hyperboloid $H_n$)
and complex projective spaces $CP^N$ have been discussed in many papers
\cite{KMP2}-\cite{NERSES}. The first investigation for one spatial dimension
was done in the two recent papers \cite{MARDO,MARDO1}. In particular
in the article \cite{MARDO} have constructed the complex mapping
$S_{1C}\to S_{1}$, which extend in spherical geometry well-known for
Euclidean space Hurwitz transformation. It was also shown that this
transformation establish the correspondence between Coulomb and
singular oscillator systems in one dimensions.

In the present note we find the relation between two exactly-solvable
potentials on one-dimensional hyperbolic space $H_1$: $s_0^2 - s_1^2=R^2$
\begin{eqnarray}
\label{POT-SOC1}
V^{so}({\vec s})
=  \frac{\omega^2 R^2}{2}
\frac{s_1^2}{s_0^2} + \frac{1}{2} \frac{k^2-\frac14}{s_1^2},
\qquad
V^{sc}({\vec s})
=  - \frac{\mu}{R}
\frac{s_0}{|s_1|} + \frac{1}{2}\frac{p^2-\frac14}{s_1^2},
\end{eqnarray}
where $s_0, s_1$ are Cartesian coordinates in the ambient
pseudo-euclidean space $E_{1,1}$ and $k$, and $p$ are positive.

The potentials in (\ref{POT-SOC1}) belong to the well-known class
of superintegrable potentials in the constant curvature spaces
\cite{GROP,KMP1} restricted to one spatial dimension.
The ordinary 1D oscillator system ($k=1/2$) is singular
at the point $s_0 = 0$, which separate the motion on the upper and
lower sheets of hyperbola. The ordinary 1D Coulomb system ($p=1/2$)
is attractive on upper and repulsive on lower sheets of hyperbola.
Below we will consider only the motion on the upper sheet of hyperbola
($s_0 > 0$).

\vspace{0.2cm}
\noindent
{\bf 2.} The Schr\"odinger equation describing the nonrelativistic
quantum motion on one-dimensional hyperbolic space $H_1$
in pseudospherical coordinates $\tau \in (- \infty, \infty)$
\begin{eqnarray}
\label{COOR1}
s_0 = R\cosh\tau,  \qquad s_1 = R\sinh\tau,
\end{eqnarray}
has the following form ($\hbar=\mu=1$)
\begin{eqnarray}
\label{SCH22}
\frac{d^2 \Psi}{d\tau^2} +  2R^2
\left[E - V(\tau) \right]\, \Psi = 0.
\end{eqnarray}
Let us consider first the eq. (\ref{SCH22}) for potential
$V(\tau) = V^{so}(\tau)$. Taking into account that in pseudospherical
coordinates (\ref{COOR1})
\begin{eqnarray}
\label{POTEN11}
V^{so}(\tau) =  \frac{\omega^2 R^2}{2}
\frac{s_1^2}{s_0^2} + \frac{1}{2}\frac{k^2-\frac14}{s_1^2}
=
\frac{1}{2R^2}\left(-\frac{k_0^2-\frac14}{\cosh^2\tau} +
\frac{k^2-\frac14}{\sinh^2\tau}\right) + \frac{\omega^2 R^2}{2},
\end{eqnarray}
where $k_0^2= \omega^2 R^4 + \frac14$ we obtain from eq.(\ref{SCH22})
\begin{eqnarray}
\label{OSCIL11}
\frac{d^2 \Psi}{d\tau^2} + \left[\epsilon +
\frac{k_0^2-\frac14}{\cosh^2\tau} -
\frac{k^2-\frac14}{\sinh^2{\tau}}\right]\, \Psi = 0
\end{eqnarray}
with $\epsilon = 2R^2E - \omega^2 R^4$. Thus the singular oscillator
problem on the hyperbola $H_1$ is described by the well-known
{\it modified  P\"oschl-Teller} equation and, unlike the Schr\"odinger
equation for 1D singular oscillator on the flat space $E_1$ and the sphere
$S_1$, the eq. (\ref{OSCIL11}) possesses both discrete ($\epsilon < 0$)
and continuous ($\epsilon \geq 0$) spectrum.

The bound state wave functions is given by \cite{GROP,KMP1}
\begin{eqnarray}
\label{EPSILON201}
\Psi_{n}^{(\pm k, k_0)}(\tau)
=
N_{n}^{(\pm k, k_0)}
\,
(\sinh\tau)^{\frac{1}{2}\pm k}
\,
(\cosh\tau)^{2n-k_0+\frac12}
\,
{_2F_1}(-n, k_0-n; \, 1 \pm k; \, \tanh^2\tau),
\end{eqnarray}
where $N_{n}^{(\pm k, k_0)}$ is the normalization constant.
The $\epsilon$ is quantized as
\begin{eqnarray}
\label{EPSILON21}
\epsilon = - (2n + 1 \pm k- k_0)^2
\end{eqnarray}
with $n=0,1,...[\frac12(k_0 \mp k-1)]$ and a bound state solution
is possible only for $k_0 \mp k > 1$. Then the discrete energy spectrum
takes the value
\begin{eqnarray}
\label{ENER21}
E_{n}^{\pm k}(R) = - \frac{1}{2R^2}\left[(2n+1 \pm k)^2 -
2k_0(2n + 1 \pm k) + \frac14\right].
\end{eqnarray}
The wave functions $\Psi_{n}^{(\pm k, k_0)}(\tau)$ satisfies the
orthogonality relation
\begin{eqnarray}
R \, \int_{0}^{\infty}
\Psi_{n}^{(\pm k, k_0)}(\tau) \Psi_{n'}^{(\pm k, k_0)}(\tau)
\, d\tau = \frac12 \delta_{n n'}
\end{eqnarray}
and the normalization factor has the form
\begin{eqnarray}
N_{n}^{(\pm k, k_0)}(R)
=
\frac{1}{\Gamma(1 \pm k)}
\,
\sqrt{\frac{(-2n+k_0 \mp k-1)\Gamma(-n+k_0)\Gamma(n+1 \pm k)}
{R \Gamma(-n+k_0 \mp k) n!}}
\end{eqnarray}
The positive sign at the $k$ has to taken whenever $k>\frac12$,
i.e. the additional term to oscillator potential is repulsive at
the origin and the motion take place only in one of the domains
$\tau \in [0, \infty)$ or $\tau \in [0, - \infty)$.

If $0< k \leq \frac12$, i.e. the additional term is attractive at
the origin, the motion take place in  $\tau \in (-\infty, \infty)$
and both the positive and negative sign must be taken into account
in the solution. Thus there exists two family of solutions go over
into each other under the change $k \to - k$. This is indicated by
the notion $\pm k$, in the formulas.

In the contraction limit $R\to \infty$, $\tau \to 0$,
$R \tau \sim x$ - fixed and $k_0 \sim \omega R^2$ we see that
the continuous spectrum is vanishing while the discrete spectrum
is infinite. It is easy to reproduce the energy spectrum
\begin{eqnarray}
\label{ENER201}
E_n^{\pm k} =
\lim_{R\to\infty} E_{n}^{\pm k}(R) =
\omega (2n + 1 \pm k)
\end{eqnarray}
and the wave function
$$
\Psi_{n}^{\pm k}(x)
=
\lim_{\scriptstyle R\rightarrow\infty\atop\scriptstyle
\tau\rightarrow 0}
\Psi_{n}^{(\pm k, k_0)}(\tau)
=
\sqrt{\frac{\sqrt{\omega}\Gamma(n+1 \pm k)}
{n! [\Gamma(1 \pm k)]^2}}
\,
(\sqrt{\omega} x)^{\frac{1}{2}\pm k}
\,
e^{-\frac{\omega x^2}{2}}
\,
{_1F_1}(-n; \, 1 \pm k; \, \omega x^2),
$$
for the flat 1D singular oscillator system \cite{HAKOB}.

\vspace{0.3cm}
\noindent
{\bf 3.} The 1D singular Coulomb potential in the pseudospherical
coordinates (\ref{COOR1}) has the form
\begin{eqnarray}
\label{COULOMB1}
V^{sc}(\tau) = - \frac{\mu}{R}
\left(\frac{s_0}{|s_1|} - 1\right) +
\frac{1}{2}\frac{p^2-\frac14}{s_1^2}
= - \frac{\mu}{R}(\coth|\tau| -1) +
\frac{1}{2R^2} \frac{p^2-\frac14}{\sinh^2\tau}.
\end{eqnarray}
Let us first consider the region $\tau \geq 0$.
Substituting the potential (\ref{COULOMB1}) in eq.(\ref{SCH22}),
we arrive to equation
\begin{eqnarray}
\label{COOS21}
\frac{d^2 \Psi}{d\tau^2} + \left[(2R^2 E - 2\mu R)
+ 2\mu R \coth\tau  - \frac{p^2-\frac14}{\sinh^2\tau}
\right]\, \Psi = 0,
\end{eqnarray}
which is  known as the {\it Manning-Rosen} potential problem \cite{MR}.
Making now the transformation from variable  $\tau$
to the new variable $\alpha \in [0, \infty)$
\begin{eqnarray}
\label{TRANS21}
e^{\tau} = \cosh\alpha,
\end{eqnarray}
and setting  $\Psi(\alpha) = W(\alpha)/\sqrt{\coth\alpha}$. As result
we obtain the equation
\begin{eqnarray}
\label{COOS22}
\frac{d^2 W}{d\alpha^2} + \left[2R^2 E +
\frac{(-{2R^2 E}+ 4\mu R) - \frac{1}{4}}{\cosh^2\alpha} -
\frac{{4p^2-\frac14}}{\sinh^2\alpha} \right]\, W = 0.
\end{eqnarray}
It is easy to see that, up to the substitution
\begin{eqnarray}
\label{OSC-COUL2}
\epsilon = {2R^2 E}, \qquad
k_0^2 = - {2R^2 E} + 4\mu R, \qquad
k^2 = 4p^2.
\end{eqnarray}
the equation (\ref{COOS22}) for the 1D singular Coulomb problem coincides
with the 1D singular oscillator equation (\ref{OSCIL11}).  The regular for
$\alpha\in [0,\infty)$ solution of this equation according to
(\ref{EPSILON201}) is
\begin{eqnarray}
\label{OSCFUN22}
W_n(\alpha) =
A_{n}
\,
(\sinh\alpha)^{\frac12 \pm k}
\,
(\cosh\alpha)^{2n-k_0+ \frac12}
\,
{_2F_1}\left(-n, -n+k_0; \, 1 \pm k; \, \tanh^2\alpha\right)
\end{eqnarray}
where $A_{n}$ is a normalization constant. The constant $A_{n}$ is
computed from the requirement that the wave function (\ref{OSCFUN22})
satisfies the normalized condition
\begin{eqnarray}
R\,
\int_{0}^{\infty}
|\Psi_{n}(\tau)|^2 \, d\tau =
R\,
\int_{0}^{\infty}
|W_{n}(\alpha)|^2 \, \tanh^2\alpha \, d\alpha = \frac12,
\end{eqnarray}
and has the following form
\begin{eqnarray}
\label{COUL-CONST21}
A_{n} =
\sqrt{\frac{k_0(k_0-2n \mp k-1)\Gamma(k_0-n) \Gamma(n + 1 \pm k)}
{R (2n+1 \pm k)(n)![\Gamma(1\pm k)]^2\Gamma(k_0-n \mp k)}}.
\end{eqnarray}
Let us now consider the two most interesting cases.

\vspace{0.3cm}
\noindent
{\bf 3.1} In the case when $p = \frac12$ the centrifugal term in
$V^{sc}$ disappears and the motion take place in the region
$\tau \in (-\infty, \infty)$. Then the transformation (\ref{TRANS21})
establishes the connection between ordinary 1D
Coulomb problem and 1D singular oscillator with $k=1$.
Comparing now  eq.(\ref{OSC-COUL2}) with (\ref{EPSILON21}), we get
\begin{eqnarray}
k_0 = (n+1+\sigma), \qquad
\sigma = \frac{\mu R}{n+1}.
\end{eqnarray}
and for the discrete energy spectrum
\begin{eqnarray}
\label{BOUND012}
E_n(R) = - \frac{(n+1)^2}{2R^2} -
\frac{\mu^2}{2(n+1)^2} + \frac{\mu}{R},
\qquad n=0,1,2,... [\sigma-1]
\end{eqnarray}
Returning to the variable $\tau$, we find for bound state wave
function ($\tau >0$)
\begin{eqnarray}
\label{COUL-HYPER1}
\Psi_{n \sigma} (\tau)
= A_n(\sigma) \,
\sinh\tau \, e^{(n-\sigma)\tau}
\,
{_2F_1}\left(-n, 1+\sigma; \, 2; \, 1-e^{-2\tau}\right),
\end{eqnarray}
where the normalization constant $A_n(\sigma)$ is
\begin{eqnarray}
\label{CONSTANT1}
A_n(\sigma) =
\sqrt{\frac{2\sigma [\sigma^2- (n+1)^2]}{R}}
\end{eqnarray}
In the domain $-\infty < \tau < 0$ 1D Coulomb wave function are
determined from eq. (\ref{COUL-HYPER1}) after reflection $\tau \to -\tau$.
Finally the general solution of the Schr\"odinger equation for
$-\infty < \tau < \infty$ then can be presented in form of even and
odd functions
\begin{eqnarray}
\label{COUL-HYPER01}
\Psi_{n \sigma}^{(+)} (\tau)
&=& A_n (\sigma) \,
\sinh|\tau| \, e^{(n-\sigma)|\tau|}
\,
{_2F_1}\left(-n, 1+\sigma; \, 2; \, 1-e^{-2|\tau|}\right),
\\[2mm]
\label{COUL-HYPER02}
\Psi_{n \sigma}^{(-)} (\tau)
&=& A_n (\sigma) \,
\sinh\tau \, e^{(n-\sigma)|\tau|}
\,
{_2F_1}\left(-n, 1+\sigma; \, 2; \, 1-e^{-2|\tau|}\right),
\end{eqnarray}
Thus we have seen that using the transformation (\ref{TRANS21})
we can completely solve 1D Coulomb system on hyperbola $H_1$, including
eigenfunctions and discrete energy spectrum.

Now let us consider the flat space contraction. In the contraction
limit $R\to\infty$ the energy spectrum for finite $n$ goes to the
energy spectrum of the 1D hydrogen atom \cite{POGTER}
\bea
\label{SS22-11}
E_n = \lim_{R\rightarrow\infty} E_n^{\nu}(R)
= - \frac{\mu^2}{2(n+1)^2},
\qquad n=0,1,2,...
\eea
In the limit $R\to \infty$, putting $\tau \sim x/R$ in eqs.
(\ref{COUL-HYPER01}) and (\ref{COUL-HYPER02}), we obtain
(up to the factor $\sqrt{2}$) the well-known 1D Coulomb wave
function \cite{POGTER}
\bea
\lim_{\scriptstyle R\rightarrow\infty\atop\scriptstyle
\tau\rightarrow 0}
\Psi_{n\sigma}^{(+)}(\varphi)
&=&
\sqrt{\frac{2\mu^3}{(n+1)^3}}\,
|x|\,
e^{-\frac{\mu |x|}{n}}\,
{_1F_1}\left(-n; \, 2; \, \frac{2\mu |x|}{n}\right),
\nonumber\\[2mm]
\lim_{\scriptstyle R\rightarrow\infty\atop\scriptstyle
\tau\rightarrow 0}
\Psi_{n\sigma}^{(-)}(\varphi)
&=&
\sqrt{\frac{2\mu^3}{(n+1)^3}}\,
x\,
e^{-\frac{\mu |x|}{n}}\,
{_1F_1}\left(-n; \, 2; \, \frac{2\mu |x|}{n}\right).
\nonumber\eea

\vspace{0.3cm}
\noindent
{\bf 3.2.} Let us now choose $k = {\frac12}^{-}$. Then
$p^2 - \frac{1}{4}= -3/16$ and the centrifugal term is
attractive at the origin for potential $V^{sc}$.
As in the previous case the motion take place in the domain
$-\infty < \tau < \infty$. For potential $V^{so}$ this term
vanish and therefore transformation (\ref{TRANS21})
connect the ordinary oscillator with 1D singular Coulomb
system.

Let us now introduce the quantity $\nu=\frac12(1\pm k)$, which
takes two values $\nu = \frac14$ and $\nu=\frac34$.
Making the all calculation by analogy to previous case, it easy to
get the discrete energy spectrum
\begin{eqnarray}
\label{ENERGY1-0}
E_n^{\nu} (R) = - \frac{(n + \nu)^2}{2R^2} - \frac{\mu^2}{2(n+\nu)^2}
+ \frac{\mu}{R},
\qquad
n= 0,1,2,....
\end{eqnarray}
and normalized bound state wave function
\bea
\label{SS22-0}
\Psi_{n \sigma_{\nu}}^{\nu}(\tau)
&=&
2^{\nu}
\,
\sqrt{\frac{[\sigma^2_{\nu} - (n+\nu)^2]\Gamma(\sigma_{\nu}+\nu)
\Gamma(n+2\nu)} {2R(n+\nu) n! \Gamma(\sigma_{\nu}+1-\nu)[\Gamma(2\nu)]^2}}.
\nonumber\\[2mm]
&\times&
(\sinh\tau)^{\nu}
\,\,
e^{\tau(n-\sigma_{\nu})}
\,\, {_2F_1}(-n, \, \nu+\sigma_{\nu}; \,
2\nu; \, 1-e^{-2\tau}),
\eea
where $\sigma_{\nu} = \mu R /(n+\nu)$.

In the flat space limit $R\to \infty$, it is easy to see
\bea
\label{SS22-110}
E_n^{\nu} =
\lim_{R\rightarrow\infty} E_n^{\nu}(R)
= - \frac{\mu^2}{2(n+\nu)^2},
\qquad n=0,1,2,...
\eea
and
\bea
\label{SS22-12}
\Psi_n^{\nu}(y)=
\lim_{\scriptstyle R\rightarrow\infty\atop\scriptstyle
\tau\rightarrow 0}
\Psi_{n\sigma}^{\nu}(\vphi)
=
\frac{\sqrt{\mu}}{\Gamma(2\nu)}
\frac{1}{(n+\nu)}
\,
\sqrt{\frac{\Gamma(n+2\nu)}{2n!}}\,
y^{\nu}\,
e^{- \frac{y}{2}}\,
{_1F_1}(-n; \, 2\nu; \, y),
\eea
where $y = {2\mu x}/{(n+\nu)}$. Formulas (\ref{SS22-110}) and
(\ref{SS22-12}) coincides with the formulas for energy levels
$E_n^{\nu}$ and up to the factor $\sqrt{2}$ for wave functions
$\Psi_n^{\nu}(y)$ for the two type of one-dimensional Coulomb
anyons with $\nu=\frac14$ and $\nu=\frac34$ respectively
\cite{TER-ANT}.

\end{document}